\documentclass[12pt]{article}

\usepackage{myArticle}
\graphicspath{{../figures/}}
\usetikzlibrary{decorations.pathreplacing,arrows, arrows.meta}

\title{Subgame perfect Nash equilibrium for dynamic pricing competition with finite planning horizon}
\author[1]{Niloofar Fadavi\thanks{nfadavi@smu.edu}}

\affil[1]{Department of Industrial Engineering, Sharif University of Technology}

\begin{document}\thispagestyle{empty}
\maketitle

\begin{abstract}
Having fixed capacities, homogeneous products and price sensitive customer purchase decision are primary distinguishing characteristics of numerous revenue management systems.
Even with two or three rivals, competition is still highly fierce.
This paper
studies sub-game perfect Nash equilibrium of a price competition in an oligopoly market with
perishable assets. Sellers each has one unit of a good that cannot be replenished, and they compete
in setting prices to sell their good over a finite sales horizon. Each period, buyers desire one unit of the good and the number of buyers coming to the market in each period is random. All sellers'
prices are accessible for buyers, and search is costless. Using stochastic dynamic programming 
methods, the best response of sellers can be obtained from a one-shot price competition game
regarding remained periods and the current-time demand structure. Assuming a binary demand model, we demonstrate that the duopoly model has a unique Nash equilibrium and the oligopoly model does not reveal price dispersion with respect to a particular metric. We illustrate that, when considering a generalized demand model, the duopoly model has a unique mixed strategy Nash equilibrium while the oligopoly model has a unique symmetric mixed strategy Nash equilibrium.
\end{abstract}
\renewcommand{\set}[1]{\mathrm{#1}}
\newcommand{\actUp}{\set{U}}
\newcommand{\actDw}{\set{L}}
\newcommand{\inact}{\set{I}}
\newcommand{\act}{\set{A}}
\newcommand{\partition}{(\actDw, \actUp, \inact)}

\newcommand{\A}[0]{\mathcal{A}}
\newcommand{\I}[0]{\mathcal{I}}
\newcommand{\B}[0]{\mathcal{B}}
\newcommand{\F}[0]{\mathcal{F}}
\newcommand{\M}[0]{\mathcal{M}}
\renewcommand{\rv}{\tilde{\omega}}
\renewcommand{\rvset}{\Omega}
\renewcommand{\obs}{\omega}
\newcommand{\e}{\Delta \rho}
\newcommand{\f}{\Delta \bar{y}}

\newcommand{\hg}[1]{\textcolor{purple}{[~HG:#1~]}}
\newcommand{\nf}[1]{\textcolor{blue}{#1}}

\section{Introduction}\label{sect:intro}
The goal of revenue management is to manage the sale of a limited
quantity of a resource (such as hotel rooms, airline seats, advertising slots, etc.) to a potential set
of customers (the market).Airlines are one example of a perishable capacity provider who competes by pricing fixed capacities for sale over a limited sales horizon. Expedia and other online travel agencies gather information and display real-time flight prices from rival airlines \cite{gallego2014dynamic}. This makes it possible for customers to compare the quality and price of products and choose the best one. In a rapidly-changing technological world, company managers have become increasingly intelligent, and revenue management has become a key factor in their decision-making process.
This is because in a competitive environment, it is important for companies to have a good understanding of their competitors' potential actions in order to make suitable decisions. For example, companies may fail to sell their products
When they set a price too high because they might make decision
Regardless of their competitors' likely behavior. For this reason, there has been considerable recent
interest and research activity to understand revenue management competition.
In this paper, we consider an oligopoly stochastic model, in which sellers compete by setting prices in
each period. This model, which is an advanced version of Bertrand-Edgworth competition aims to
determine the managers’s consideration and the pricing behavior in this uncertain situation.

The remainder of the paper is structured out as follows: We review the related literature in section \ref{sec:liter}. In section \ref{sec:meth}, we describe the problem and provide equilibrium properties for different demand settings. 
Theoretical evidence for the declared properties is presented in section \ref{sec:abs}.
The results of the numerical experiments are presented in \ref{sec:num}. Finlay, the research conclusions and future research opportunities are discussed in \ref{sec:conclusion}.

\section{Related Literature} \label{sec:liter}
We have divided the related works into two categories in this section. The discussion of dynamic competition models is followed by a brief introduction of static competition models.
\subsection{Static Models}
The two classic models of oligopoly are the Cournot model in which firms compete on quantities \cite{cournot1838recherches}
and the Bertrand model in which firms compete on prices \cite{bertrand1988review}. The Cournot model starts with companies deciding how much to produce, and the market then sets the prices. According to the Bertrand model of price competition, consumers only shop at businesses who have the lowest prices because they make similar goods. At the beginning of the period, prices are set by competing businesses.
The Bertrand model also makes the assumption that each firm can meet the demand for its product as it sees it. In this model, each firm sets its price at its marginal cost, and there is only one possible equilibrium.
In his study, Edgworth demonstrated that, when capacity constraints are taken into account, there is no pure strategy Nash equilibrium in the Bertrand model \cite{edgeworth1925pure}. This study focuses on the Bertrand-Edgeworth model, which takes into account Bertrand competition with capacity constraints.
In this case, clients pick the company with the lowest price, and when two companies have comparable prices, they make a random selection. Customers arrive at the market in a random order, which introduces demand uncertainty. Although there isn't necessarily a pure-strategy equilibrium in this situation, there are mixed-strategy equilibria \cite{allen1986bertrand}, \cite{dasgupta1986existence}, \cite{levitan1972price}.
Bertrand-Edgeworth appears to be a good place to start if we want to learn more about revenue management competition. The majority of the time, revenue management is concerned with situations where a set capacity needs to be sold quickly (like the quantity of hotel rooms or airplane capacity), and businesses often compete on pricing \cite{martinez2011dynamic}.
The absence of a pure-strategy equilibrium makes it challenging to study and forecast potential market circumstances. The absence of a pure-strategy equilibrium makes it challenging to study and forecast potential market circumstances. The model (Bertrand-Edgeworth) is straightforward, however it doesn't give precise information about the state of the market.
Therefore, it would be useless to design the model for more complex cases like several client segments.
As a result, some of the other pertinent publications were concentrating on mixed-strategy Nash equilibrium at the same time as these fundamental models were being developed.
\subsection{Dynamic Models}
Many researchers have taken into consideration dynamic models of competition after the emergence of numerous static models. In this part, we focus on the most significant ones.
Chamberlin \cite{chamberlin1933theory} was of the first researchers to point out that repeated interaction between
competitors in an oligopoly market, can facilitate collusion. When businesses interact frequently enough, they can threaten price wars in order to deter rivals from engaging in collusive pricing.
Dudy’s paper is one of the earliest papers on dynamic competitions with capacity constraints  \cite{dudey1992dynamic}.
In spite of its nonexistence in Bertrand-Edgworth competition, Dody proved that subgame perfect equilibrium is unique for dynamic competition. Proving equilibrium's uniqueness was a significant achievement. This is because non-uniquness of equlibrium is problematic for
 economic applications. Without a unique result, we do not know which equilibrium is
exactly considered by each competitor, and we cannot legitimately perform comparative static
exercises \cite{osborne1986price}.
Dudy established that the sub-game perfect equilibrium is unique for dynamic competition while not being in Bertrand-Edgworth competition. The proof of the uniqueness of equilibrium was a great accomplishment. This is due to the fact that non-uniquness of equilibrium presents difficulties in economic applications. Without an unique result, we are unable to compare static exercises amongst competitors because we cannot determine precisely which equilibrium each competitor is taking into account.
In Dudy's model, companies in each period set a price to maximize their undiscounted profits in a duopoly variant of Bertrand-Edgworth competition with capacity constraint.
They assumed that the demand structure is deterministic in their model.\\
Biglaiser and Vettas \cite{biglaiser2004dynamic}, investigate a dynamic price game in which businesses have limited capacities, and customers are strategic. They discover that there is no pure-strategy subgame perfect equilibrium, and that seller's market shares are highly likely to be asymmetric.\\
The key contribution of \cite{anderson2007price} is exploring the influence of search costs on the equilibrium in a dynamic price model. They demonstrated that when the search cost is not zero, duopolistic prices are greater than those of monopolists.\\
Mantin \cite{mantin2008effect} analyzed a multiperiod duopoly pricing game where a homogeneous perishable good is
sold to consumers who visit one of the retailers in each period.\\
Perakis and Sood in \cite{perakis2004competitive}, studied an open-loop equilibrium. Prices are supposed to be set at the start of the horizon and cannot change until the end of the horizon, therefore they do not need to be subgame perfect. They used variational inequality tools and performed comparative
statics on equilibrium prices as a function of demand- capacity ratios.\\
Dynamic price dispersion in perfect competition was derived in Deneckere and Peck \cite{deneckere2012dynamic}. They
developed a dynamic version of Prescott’s hotel model. Their model presents a
price dispersion for each period, and updates the demand function regarding the last periods
revealed demand. In each period, some firms post a low price and sell for sure, while others post a higher price and can only sell during that period if demand is high. Firms update their knowledge and make decisions for future periods based on their revised demand function when demand is revealed. In their research, no discount factor was taken into account.\\
Talluri and Martínez de Albéniz in \cite{martinez2011dynamic}, studied perfect competition of a homogeneous product when
demand function is a Bernoulli random variable, and derived a closed-form solution to the equilibrium
price paths. They showed a structural property of the equilibrium policy such that the seller with the
lower equilibrium reservation value sells a unit at a price equal to the competitor’s equilibrium
reservation value.\\
\cite{gallego2014dynamic} analyzed a deterministic demand function in the first section of their work, and then proved that under certain special conditions, the first section's results can be utilized as an approximation of the circumstance where the demand function is random.\\
\cite{sun2017dynamic} considered an oligopolistic competition in which each firm has one unit to sell, and
products are not perishable. In his model, each firm tries to maximize its discounted expected profit. They
discovered the equilibrium price dispersion, and they proved it is unique when demand follows Bernoulli 
dispersion.\\
Although there are a huge number of researches done in the last decade, their concentration is mostly
on the customer behavior modeling and its enhancement. On the other hand, there is an obvious lack
of attention to competition modeling. This is because when we add a constraint to these models, it
becomes difficult to prove the existence of subgame perfect equilibrium in a stochastic revenue management game.
Even the simplest model, Bertrand-Edgworth model, does not have a unique pure Nash equilibrium,
so it seems difficult to develop these models. 
Also most of the related studies are in static models and
infinite games, and some of the studies considered special situation to find pure Nash equilibrium with
no regard to mixed strategies’ existence.
Our main contribution is that we find equilibrium for a finite long run horizon, and we show that there is a unique mixed strategy Nash equilibrium in the
duopoly model with the
generalized demand. 
Furthermore, we show that the symmetric mixed strategy Nash equilibrium is unique
for oligopoly model.
In order to be able to examine how various factors affect the state of the market and sellers' behavior in equilibrium, our goal is to create a mathematical model for the given problem. Although it is difficult to meet all of the model's assumptions, this model can be used as a benchmark to understand how the various parameters affect the market's behavior and price dispersion's features. For this aim, we make use of the mathematical model presented in  \cite{sun2017dynamic}.

\section{Methodology} \label{sec:meth}
In this section,  we examine a dynamic Bertrand-Edgworth competition. We first discuss Bernoulli demand distribution before moving on to general demand distribution. For each demand setting, We examine equilibrium for both doupoly and oligopoly competitions.

Let's first discuss the fundamentals of Nash equilibrium before delving into the specifics of the equilibrium study. 
Think of a strategic game with $N$ players. The set of pure strategies for competitor $i$ is $A_i$, then a mixed strategy is a randomization on set $A_i$ and is denoted by $F$. The mixed strategies of the present competitors are assumed to be independent. Finally, we have a new game with mixed strategies rather than the previous simple pure strategies. This game is called mixed extension of the original game. Now, we have the following definition from \cite{osborne1994course}:
\begin{definition}
 "A mixed strategy Nash equilibrium of a strategic game is a Nash equilibrium of its mixed extension."
\end{definition}
The other useful concept is subgame perfect Nash equilibrium (SPNE). This is a variation of Nash equilibrium applied to dynamic games. A set of strategies demonstrate a SPNE, if each point of this dynamic game, which is a subgame, results in a Nash equilibrium \cite{osborne1994course}. 

 \subsection{The Model and Notation}
 We consider a market that includes $N$ Competing firms, each selling one unit of a homogeneous perishable item. Time is considered as a discreet parameter and is indexed by $t=1,2, \ldots .$ in an infinite horizon. Each firm should set a price before the demand is revealed in every period. Customers only buy one unit if the price is less than their reservation price, which is the same for all of them. Once a customer arrive to the market in period $T$, they ask for one unit of the product and leave the market if they fail to buy in this period. When two businesses set their prices equally, they have an equal chance of selling their goods. Each firm's purpose is to maximize the present discounted value of its profit. To discuss the variants of this problem, let us first introduce certain notations:
\begin{itemize}
\item $ T$ : number of time units left until the product expires.
\item $\bar{p}$ : reservation Price.
\item $ V(\bar{p}, \mathrm{~N}, \mathrm{~T})$ : the expected firm's option value when there are $N$ firms in the market and $T$ periods remained.
\item $ P_{N, T}^{*}$ : the minimum price each firm tend to consider when there are $N$ firms in the market and $T$ periods remained.
\item $ \delta$ : discount factor.
\end{itemize}
Calculating option values is a crucial factor to consider when developing a pricing strategy for a company. This option value is equal to the expected profit they make if they do nothing and keep the price at the reservation price till the planning horizon is over \cite{sun2017dynamic}. Because they have this option, they have a lower bound for the profit of their pricing strategy they adopt. So, from definition of option value, we have the following equation:
\begin{align}
\mathrm{V}(\bar{p}, \mathrm{~T}, \mathrm{~N})=0
\end{align}
for $N>T$.

In the next sections, we explore Nash equilibrium for the duopoly model, then create and find the equilibrium for the oligopoly market.
\subsubsection{Binary Demand}
In this section, we study the market when the demand distribution follows Bernoulli distribution. We first discuss some properties of option value that apply to both duopoly and oligopoly markets. Following that, we discuss the equilibrium for each case individually. Let us first introduce some additional notations:
\begin{itemize}
    \item $ q$ : probability of zero demand.
    \item $\Gamma(N, q, T)$ : a game in which $\mathrm{N}$ firms are present in the market, and planning horizon is $T$, and demand parameter is equal to $q$.
\end{itemize}
We can perform the computations for our case using the same reasoning as \cite{sun2017dynamic}. For this market, closed-form of the option value can be calculated. Based on Bayes' theorem, we have:
\begin{align} \label{eq:num2}
\mathrm{V}(\bar{p}, \mathrm{~N}, \mathrm{~T})=q \delta V(\bar{p}, N, T-1)+(1-q) \delta V(\bar{p}, N-1, T-1).
\end{align}
Similarly, $\mathrm{V}(\bar{p}, 1, \mathrm{~T})$ can be computed as follows :
\begin{align} \label{eq:num3}
V(\bar{p}, 1, \mathrm{~T})=q \delta V(\bar{p}, 1, T-1)+(1-q) \bar{p}=\\ \notag
(1-q) \bar{p}+q \delta(1-q) \bar{p}+\ldots+(q \delta)^{T-1}(1-q)\bar{p}=\\ \notag
(1-q) \bar{p}(1-(q \delta)^{T} ) /(1-q \delta).
\end{align}
According to \eqref{eq:num2} and \eqref{eq:num3}, we can calculate $\mathrm{V}(\bar{p}, \mathrm{~N}, \mathrm{~T})$ according to following equation:
\begin{align}
\mathrm{V}(\bar{p}, \mathrm{~N}, \mathrm{~T})=\sum_{i=1}^{T-N+1}[(q \delta)^{i-1}(1-q) \delta V(\bar{p}, N-1, T-i)],
\end{align}
when $N \leq T$.
Because firms have an option value, there is a lower bound for the price they are willing to accept in a given period. To figure out what this price is, we need to find a price $P_{N, T}^*$ where the expected profit from posting this price is equal to the current option value. As a result, it fulfills the following equation:
\begin{align}
\mathrm{V}(\bar{p}, \mathrm{~N}, \mathrm{~T})=q \delta \mathrm{V}(\bar{p}, \mathrm{~N}, \mathrm{~T}-1)+(1-q) P_{N, T}^{*}
\end{align}
for $N \leq T$. In conclusion, having $V(\bar{p}, N, T)$, we can  calculate $P_{N, T}^{*}$ as:
\begin{align} \label{eq:num6}
P_{N, T}^{*}=\frac{\mathrm{V}(\bar{p}, \mathrm{~N}, \mathrm{~T})-q \delta \mathrm{V}(\bar{p}, \mathrm{~N}, \mathrm{~T}-1)}{1-q}=\delta \mathrm{V}(\bar{p}, \mathrm{~N}-1, \mathrm{~T}-1)
\end{align}
for $N \leq T$.
Equation \eqref{eq:num6} has an intuitive explanation as well. This idea which is mentioned in \cite{sun2017dynamic} for nonperishables assets works here as well. When the demand distribution is Bernoulli, in each period at most only one buyer comes to the market. Thus, $P_{N, T}^{*}$ is the price at which a seller is indifferent between selling or letting others sell when there is a one unit demand. The profit in the first case is $P_{N, T}^{*}$, and the profit in the second case is $\delta \mathrm{V}(\bar{p}, \mathrm{~N}-1, \mathrm{~T}-1)$. Therefore, $P_{N, T}^{*}=\delta \mathrm{V}(\bar{p}, \mathrm{~N}-1, \mathrm{~T}-1)$. 

\subsubsection{Duopoly Competition with Binary Demand} 
In this section we examine the equilibrium for duopoly markets. To examine the mixed-strategy equilibrium, we model the sellers' strategy as a random variable. Here, we introduce some notations:
\begin{itemize}
\item $ F_{iT} (p)$ : the probability that  seller $i$ sets a price less than or equal to $p$ in period $T$,
\item $ F_{-i T}(p)$ : the joint probability that all sellers except $i$ set prices less than or equal to $p$ in period
$T$,
\item $ F_{i T}(p_{-})=\lim _{p \rightarrow p^{-}} F_{i T}(p)$,
\item $ I_{i T}=\sup \{p \mid F_{i T}(p)=0\}$,
\item $ U_{i T}=\inf \{p \mid F_{i T}(p)=1\}$.
\end{itemize}
Lemma \ref{lem:one} and \ref{lem:two} demonstrate the relationship between the minimum acceptable price $P_{N, T}^{*}$ and the bounds associated with the adopted random pricing strategy. These two together help to attain the equilibrium properties in proposition \ref{lem:three}.
\begin{lemma} \label{lem:one}
The following equation holds in equilibrium:
\begin{align}
I_{1 T}=I_{2 T} \geq P_{2, T}^{*}
\end{align}
\end{lemma}
\begin{lemma} \label{lem:two}
The following equation holds in equilibrium:
\begin{align}
U_{1 T}=U_{2 T}=P_{2, T}^{*}.
\end{align}
\end{lemma}
\begin{proposition} \label{lem:three}
In the selling game $\Gamma(2, q, T)$, each seller adopts the pricing strategy as $\mathrm{p}=P_{2, T}^{*}$ for all $T \geq 2$,  and the expected profit for each seller is exactly $\delta V(\bar{p}, 1, T)$.
\end{proposition}
As a result, when there are two firms competing on price to sell their product, they both offer an identical price equal to $P_{2, T}^{*}$. The reason for this is that no one intends to sell for less than their option value, and if they let the other person sell, their expected profit is the same because of equation  \eqref{eq:num6}. 
\subsubsection{Oligopoly Competition with Binary Demand}
In this section, we examine the equilibrium when there are more than two firms in the market. First, we introduce some notations for this part:
\begin{itemize}
    \item$ F_{i N T}(p)$ : the probability that seller $i$ sets a price less than or equal to $p$ in period $T$ in a market containing $N$ sellers,
    \item$ F_{-i N T}(p)$ : the probability at least one seller except seller $i$ set a price less than or equal to $p$ in period $T$ in a market containing $N$ sellers,
    \item$ F_{i N T}(p_{-})=\lim _{p \rightarrow p^{-}} F_{i N T}(p)$,
    \item$ \gamma_{N T}(p)$ : the number of firms offering a price equal to $p$,
     \item$ I_{i N T}=\sup \{p \mid F_{i N T}(p)=0\}$,
     \item$ U_{i N T}=\inf \{p \mid F_{i N T}(p)=1\}$.
\end{itemize}
Now, we can discuss the equilibrium properties of this oligopoly market. 
\begin{proposition} \label{lem:four}
When there are N firms in the market and the demand structure is binary, the three following conditions hold in equilibrium:\\
1. $I_{i N T} \geq P_{N, T}^{*}$.\\
2. $\gamma_{N T}\left(p_{N, T}^{*}\right) \geq 2$.\\
3. The expected profit for each firm is $V(\bar{P}, N, T)$.
\end{proposition}
Based on the proposition, when the demand structure is binary, offering a price more than the reservation price is possible, and the price dispersion is not predictable. However, based on the measure presented in \cite{baye2004price}, and the corresponding case discussed in \cite{sun2017dynamic}, if we only consider marginal prices, this oligopoly model does not show any price dispersion.

\subsubsection{General Demand Model}
In this section, we have a generalization of the cases that we discussed before. As a result, we investigate a general demand structure and investigate the duopoly and oligopoly models' equilibria. We start by introducing a few extra notations:
\begin{itemize}
  \item $q_{i}$: the probability that the demand equals to $i$.
  \item $F(p)$ : the identical assumed strategy chosen by all of the firms.
    \item $F_{iNT}(p)$ : the probability that seller $i$ sets a price less than or equal to $p$ in period $T$ in a market containing $N$ sellers.
    \item$ F_{-i N T}(p)$ : the probability that at least one seller except $i$ set a price less than or equal to $p$ in period $T$ in a market containing $N$ sellers.
    \item $ F_{i N T}\left(p_{-}\right)=\lim _{p \rightarrow p^{-}} F_{\text {iNT }}(p)$.
    \item $ I_{i T}=\sup \left\{p \mid F_{i T}(p)=0\right\}$.
    \item $ U_{i T}=\inf \left\{p \mid F_{i T}(p)=1\right\}$.
\end{itemize}
Now, similar to the binary demand section, we first calculate the option value and reservation price. The option value can be calculated from the following equation:
\begin{align} \label{eq:optionofgen}
V(\bar{p}, N, T)=\sum_{i=0}^{N-1} q_{i} \delta V(\bar{p}, N-i, T-1)+\sum_{i=N}^{+\infty} q_{i} \bar{p}.
\end{align}
The first component of \eqref{eq:optionofgen} is the expected profit we get if the demand is less than the number of players, and the second one represents the expected profit of the cases where demand is more than the number of players.
In addition, since $P_{N, T}^{*}$ is the least price that sellers are willing to offer, it satisfies the following equations:
\begin{align} \label{eq:optionofgen2}
V(\bar{p}, \mathrm{~N}, \mathrm{~T})=q_{0} \delta V(\bar{p}, \mathrm{~N}, \mathrm{~T}-1)+\left(1-q_{0}\right) P_{N, T}^{*}.
\end{align}
So, we have:
\begin{align} \label{eq:pnt}
P_{N, T}^{*}=\frac{V(\bar{p}, \mathrm{~N}, \mathrm{~T})-q_{0} \delta V(\bar{p}, \mathrm{~N}, \mathrm{~T}-1)}{1-q_{0}}=\frac{\sum_{i=1}^{N-1} q_{i} \delta V\left(\bar{p}, \mathrm{~N}-\mathrm{i}_{1}, \mathrm{~T}-1\right)+\sum_{i=N}^{+\infty} q_{t} \bar{p}}{1-q_{0}}
\end{align}
Equation \eqref{eq:pnt} has an intuitive explanation similar to the one mentioned by \cite{sun2017dynamic} for the perishable assets. It means that if demand is not zero, $P_{N,T}^{*} $ is equal to the expected option value.\\
The last parameter that can be calculated regardless of the demand distribution is the monopolist's option value, and it satisfies the following equation:
\begin{align} \label{eq:recursivgen}
V(\bar{p}, 1, T)=q_{0} \delta V(\bar{p}, 1, T-1)+\left(1-q_{0}\right) \bar{p}.
\end{align}
Using the equation \eqref{eq:recursivgen}, a closed-form of $V(\bar{p}, 1, T)$ can be derived as follows:
\begin{align} \label{eq:monopolyopt}
&V(\bar{p}, 1, T)=\left(1-q_{0}\right) \bar{p}+q_{0} \delta\left(1-q_{0}\right) \bar{p}+\ldots+q_{0} \delta^{T-1}\left(1-q_{0}\right) \bar{p} \\
&=\left(1-q_{0}\right) \bar{p} \frac{1-\left(q_{0} \delta\right)^{T}}{1-q_{0} \delta}. \notag
\end{align}
In the following sections, we first analyse the equilibrium for duopoly and then for oligopoly market.
\subsubsection{Duopoly Competition with General Demand Function}
To examine the properties of equilibrium, we need to first calculate the option value. From the option value we obtained for monopolist, the option value for a firm in a duopoly market can be calculated from the following equation:
\begin{align} \label{eq:fiftheen}
V(\bar{p}, 2,1)=\left(1-q_{0}-q_{1}\right) \bar{p},
\end{align}
and we have:
\begin{align} \label{eq:sixtheen}
V(\bar{p}, 2, T)=\sum_{l=0}^{1} q_{i} \delta V(\bar{p}, 2-\mathrm{i}, \mathrm{T}-1)+\sum_{i=2}^{+\infty} q_{i} \bar{p}
\end{align}
for $T>2$.
The right-hand side of equation \eqref{eq:sixtheen} can be calculated using the option value of $T=1$ and the monopolist option value in \eqref{eq:monopolyopt}. The following proposition introduces the symmetric mixed strategy equilibrium for the duopoly case. Similar result exists and is proved in \cite{sun2017dynamic} for not perishable assets. The proof of this proposition is stated in section \ref{sec:abs}.
\begin{proposition} \label{lem:five}
Considering $q_{1}>0$ and $\sum_{l=2}^{+\infty} q_{l}>0$, the mixed strategy $W_{T}(p)$ represents a Nash equilibrium, and its associated expected profit is $V(\bar{p}, N, T)$.
\begin{align}
W_{T}(p)=\left\{\begin{array}{cc}
0 & p \leq P_{N, T}^{*} \\
\frac{V(\bar{p}, 2, T)-q_{0} \delta V(\bar{p}, 2, T-1)-\sum_{i=1}^{+\infty} q_{i} p}{q_{1}(\delta V(\bar{p}, 1, T-1)-p)} & P_{N, T}^{*}<p<\bar{p} \\
1 & p \geq \bar{p}
\end{array}\right.
\end{align}
\end{proposition}
In proposition \ref{lem:five}, we show that $W_{T}(p)$ can result in an equilibrium. Another characteristic we discovered for the duopoly case is the equilibrium's uniqueness. To show that this is a unique equilibrium, we take advantage of the expected profit in equilibrium and the equilibrium strategy's. The evidence we need to demonstrate the uniqueness in \ref{lem:eight} can be found in the following lemmas.
\begin{lemma} \label{lem:six}
In equilibrium the following equations hold.
\begin{align}
&\checkmark I_{1 T}=I_{2 T}=P_{2,T}^{*} \\
&\checkmark U_{1 T}=U_{2 T}=\bar{P}^{*}
\end{align}
\end{lemma}
\begin{lemma} \label{lem:seven}
The equilibrium strategy is symmetric.
\end{lemma}
\begin{proposition} \label{lem:eight}
$W_{T}(p)$ is the unique mixed-strategy equilibrium.
\end{proposition}
\subsubsection{Oligopoly Competition with General Demand Function}
We define the function $Z_{K, N}(x)$ according to the following equation:
\begin{align} \label{eq:twenteen}
Z_{K, N}(x)=\sum_{i=0}^{K}\left(\begin{array}{c}
N-1 \\
i
\end{array}\right)(1-x)^{N-l-1} x^{l}.
\end{align}
If all of the firms choose an identical strategy $F(p)$, the probability of an observation in which at most $k$ firms offer a price less than or equal to $\mathrm{p}$ is $Z_{K, N}(F(p))$.\\
Same as duopoly case, we can again calculate the option value as follows:
\begin{align}
V(\bar{p}, N, 1)=\sum_{l=N}^{+\infty} q_{l} \bar{p}.
\end{align}
Thus, having $V(\bar{p}, N, 1)$ and monopolist option value, the values of $ P_{N, T}^{*}$ and $V(\bar{p}, N, 1)$ can be calculated. Having the equilibrium for oligopoly cases in \cite{sun2017dynamic} for not perishable assets, the following statement proposes a symmetric mixed strategy equilibrium.
\begin{proposition} \label{lem:11}
With the conditions $\sum_{i=2}^{+\infty} q_{i}>0$, when every firm choose $W_{N,T}(p)$ as its strategy, this condition results in a unique symmetric equilibrium and their expected profit in equilibrium is $V(\bar{P}, N,T)$:
\begin{align}
&W_{N, T}(p)=\left\{\begin{array}{cc}
0 & p \leq P_{N, T}^{*} \\
G^{-1}(p) & P_{N, T}^{*}<p<\bar{p} \\
1 & p \geq \bar{p}
\end{array}\right. \\
&G(x)=\frac{\sum_{l=1}^{+\infty} q_{i} \bar{p}+\sum_{l=1}^{n-1} q_{i} z_{i-1}(x) \delta V(\bar{P}, N-i, T-1)}{\sum_{l=1}^{+\infty} q_{i}+\sum_{i=1}^{n-1} q_{i} Z_{i-1}(x)}
\end{align}
\end{proposition}

As it was mentioned before, it is important to examine uniquness of equilibrium. Therefore, the following lemma is employed in proposition \ref{lem:13} to demonstrate the uniqueness of equilibrium under some particular circumstances. For this aim, we need to provide a situation in which asymmetric equilibrium is not possible because we showed that the only possible symmetric strategy is $W_{N, T}(p)$.
\begin{lemma} \label{lem:12}
With the conditions $\sum_{i=2}^{+\infty} q_{l}>0, W_{T}(p)$ is a right continuous increasing function at $p \in$ $\left(P_{N, T}^{*}, \bar{p}\right)$.
\end{lemma}
\begin{proposition} \label{lem:13}
If the conditions $\sum_{l=2}^{+\infty} q_{i}>0$ and \eqref{eq:lowerbound23} are satisfied, the equilibrium should be symmetric.
\begin{align} \label{eq:lowerbound23}
p_{N, T}^{*} \leq I_{1 N T}=l_{2 N T}=\cdots=l_{N N T}
\end{align}
\end{proposition}
\section{Appendix} \label{sec:abs}
For the sake of proof following, we first define a few notations.
\begin{enumerate}
\item Duopoly competition with binary demand: 
The equilibrium expected value of the profit of firm $i$ is denoted by $\bar{V}_{i}^2(q,T)$. In addition the expected value of the profit of firm $i$ when firm $j$'s strategy $F_{j T}$ is given, is indicated by $V_{i}^2(p;q,T)$ and is calculated as follows:
    \begin{align} \label{eq:defvij2}
        V_{i}^2(p;q,T) &= q \delta \bar{V}_{i}^2(q,T-1) 
        +(1-q)[F_{j T}(p_{-})  V(\bar{P}, 1, T-1)\\
        &+(F_{j T}(p)-F_{j T}(p_{-})) \frac{1}{2}(\delta V(\bar{P}, 1, T-1)+p) \\&+(1-F_{j T}(p)) p] \notag
    \end{align}
\item Oligopoly competition with binary demand: The equilibrium expected value of the profit of firm $i$ is denoted by $\bar{V}_{i}^N(q,T)$. The expected value of the profit of firm $i$ when we are given $F_{-i N T}$ is given, is indicated by $V_{i}^N(p;q,T)$ and is calculated as follows:
    \begin{align} \label{eq:defvn}
        V_{i}^N(p;q,T)&=q \delta \bar{V}_{i}^N(q,T-1)
        +(1-q)\left[F_{-i N T}(p_{-}\right) \delta V(\bar{P}, N-1, T-1)\\
        &+(F_{-i N T}(p)-F_{-i N T}(p_{-}))(\frac{m}{m+1} \delta  \bar{V}_{i}^{N-1}(q,T-1)+\frac{1}{m+1} p) \\
        &+(1-F_{-i N T}(p)) p]. \notag
    \end{align}
Parameter $m$ can be calculated using the joint distribution of all $N-1$ firms. For out purpose, it is enough to know that it is a scalar in $[1,N-1]$.
\item Duopoly competition with general demand: The expected value of the profit of firm $i$ when firm $j$'s strategy $F_{j T}$ is given, is indicated by $W_{i}^2(p;q,T)$ and is calculated as follows:
    \begin{align} \label{eq:DuopolyGEXP}
         W_{i}^2(p;q,T)& = q_{0} \delta \bar{W}_{i}^2(q,T-1) 
        +q_{1}[F_{j T}(p_{-}) \delta V(\bar{P}_{,},1, T-1)\\
        &+(F_{j T}(p)-F_{j T}(p_{-})) \frac{1}{2}(V(\bar{P},1, T-1)+p)+(1-F_{j T}(p)) p] \\
        &+\sum_{i=2}^{+\infty} q_{i}p,
    \end{align}
\end{enumerate}
Here, we introduce a crucial theory that is applied throughout the proofs from \cite{osborne1994course}.
\begin{theorem} \label{lem:4.1}
If we have game with infinite strategy set, In this case, $(F_1,\ldots,F_N)$ is a mixed strategy Nash equilibrium of if and only if for every player $i$ no action in its strategy set yields, given $\{F_1,\ldots,F_N\}\setminus F_i$ , a payoff to player $i$ that exceeds his equilibrium payoff, and the set of actions that yield, given $\{F_1,\ldots,F_N\}\setminus F_i$, a payoff less than his equilibrium payoff has probability zero.
\end{theorem}
Let's now discuss the proofs offered for the equilibrium properties. Since several equilibrium characteristics are comparable in both the finite and infinite cases, we might frequently mimic the infinite case proofs provided in \cite{sun2017dynamic} to show the same characteristics in our context.

\textbf{A.1 Proof of lemma \ref{lem:one}.} 
\begin{proof}
Let us use proof by contradiction. First, we assume $I_{1 T}<I_{2 T}$.
Considering this assumption, firm $1$ can increase its profit by assigning the probability of the interval $\left(I_{1 T}, I_{2 T}\right)$ to a point in the left neighborhood of $I_{2 T}$. 
The reason is that they are sure that all of these prices would be sold with probability one if there exists any demands, thus they try to choose the maximum of them.
So, this is a contradiction arising from our false assumption, and we have $I_{1 T} = I_{2 T}$ in equilibrium. 
In addition, if we assume $I_{1 T} = I_{2 T} < P_{2, T}$ in equilibrium, there will be two possible states, and we demonstrate that neither of them can lead to equilibrium. Therefore, we prove that the equation $I_{1 T}=I_{2 T}<P_{2,T}$ cannot stand in equilibrium.
\begin{enumerate}
    \item Assume $F_{2T}(I_{2T})<1$. Because cumulative distribution function is right-continuous, we have:
    \begin{align}
        \exists \eta \in\left(l_{2T}, P_{2, T}^{*}\right) \mid \forall p \in[I_{2T}, \eta) \rightarrow F_{2 T}(I_{2T}) \leq F_{2 T}(p)<1.
    \end{align}
Now, considering firm $2$'s strategy, firm $1$ achieves smaller profit for $p \in\left[I_{2T}, \eta\right)$ than deciding not to sell in this period. The following equation demonstrates this. Utilising equation \eqref{eq:num6} and $p < P_{2, T}^{*} = \delta V(\bar{P}, 1, T-1) $, we have:
    \begin{align} \label{eq:5.2}
        V_{1}^2(p;q,T) < q \delta \bar{V}_{1}^2(q,T-1) + (1-q) P_{2, T}^{*}.
    \end{align}
The inequality \eqref{eq:5.2} is obtained by substituting $p$ by its upper bound $\delta V(\bar{P}, 1, T-1)$ in \eqref{eq:defvij2}, and the right-hand side demonstrates the expected profit of firm $1$, when they allow the other seller to sell.
As a result, based on theorem \ref{lem:4.1}, it is not possible to have a pure strategy with better result than the equilibrium's one. So, firm $1$ never offers $p \in [I_{2 T}, \eta)$, and this is a contradiction. So, state 1 is not possible in equilibrium.
\item Assume $F_{2T}\left(I_{2T}\right)=1$ in equilibrium. The, firm $2$'s expected profit associated with $p=I_{2 T}$ is absolutely less than its corresponding value for postponing the selling. This is shown in the following inequality:
\begin{align} \label{eq:Ileqp}
    V_{2}^2(p;q,T) < q \delta \bar{V}_{2}^2(q,T-1) + (1-q)P_{2,T}^{*}.
\end{align}
The right-hand side of \eqref{eq:Ileqp} is written based on the assumption $I_{2 T} < p_{2, T}^{*}$ and by substituting $p$ in \eqref{eq:defvij2} with its upper bound $P_{2, T}^{*}=\delta V(\bar{P}, 1, T-1)$. Because firm $2$ has the option to increase its profit by choosing not to sell, this state cannot be an equilibrium according to the definition of a Nash equilibrium.
\end{enumerate}
\end{proof}

\textbf{A.2 Proof of lemma \ref{lem:two}.}
\begin{proof}
First, we assume $U_{1 T} \neq U_{2 T}$ and $U_{1 T}<U_{2 T}$. Then, from lemma \ref{lem:one}, we have $p_{2, T}^{*} \leq U_{1 T}<U_{2 T}$. 
If $U_{1 T} < U_{2 T}$ in equilibrium, there will be four possible states, and we demonstrate that none of them can lead to equilibrium. In other words, we prove that the equation $U_{1 T} \neq U_{2 T}$ is not possible in equilibrium.
\begin{enumerate}
    \item $P_{2, T}^* < U_{1 T} < U_{2 T}$:
     Considering $p_1 \in\left(P_{2, T}^{*}, U_{1 T}\right)$ and $p_2 \in(U_{1 T},U_{2 T})$ the following equation holds:
        \begin{align} \label{eq:stae1}
             V_{2}^2(p_1;q,T)  > q \delta \bar{V}_{2}^2(q,T-1) + (1-q) \delta V(\bar{P}, 1, T-1) = V_{2}^2(p_2;q,T) .
        \end{align}
    The inequality is obtained from substitution of $p_1$ with its lower bound $P_{2,T}^* = \delta V(\bar{P}, 1, T-1)$ in the function $V_{2}^2(p_1;q,T)$.
     Therefore, firm $2$'s expected profit for $(U_{1 T},U_{2 T})$ is less than the equilibrium profit, and based on theorem \ref{lem:4.1} this interval should be assigned zero probability. This is in contradiction with our assumption $P_{2,T}^{*}<U_{1 T}<U_{2 T}$.
     \item $P_{2,T}^*=U_{1 T}<U_{2 T}$: From lemma \ref{lem:one}, we can conclude $P_{2,T}^{*} = I_{1 T} = U_{1 T} = I_{2 T}$. For $p_1 \in( P_{2, T}^*, U_{2 T})$, the following inequality holds:
        \begin{align} \label{eq:state2lem2}
             V_{1}^2(p_1;q,T)
             > q \delta \bar{V}_{1}^2(q,T-1)+(1-q) \delta V(\bar{P}, 1, T-1) = V_{1}^2(P_{2, T}^*;q,T).
        \end{align}
    We have \eqref{eq:state2lem2} by substituting $p_1$ with its lower bound $P_{2, T}^* = \delta V(\bar{P}, 1, T-1)$ in \eqref{eq:defvij2}.
    Therefore, firm $1$'s profit for $p_1 \in\left(p_{2,T}^*, U_{2 T}\right)$ is greater than its profit for $p_{2,T}^{*}$. Thus, it is in contradiction with the definition of Nash equilibrium, and this state cannot be an equilibrium.

    As a result of rejection of state 1 and 2, we can conclude that the equation $U_{1 T}=U_{2 T}$ should stand in equilibrium.
    \item Assume $U_{1 T} = U_{2 T} > I_{1 T} = I_{2 T} = P_{2, T}^{*}$. Because cumulative distribution function is right-continues, we have:
    \begin{align}
        \exists \eta \in (P_{2, T}^{*}, U_{2 T}) \mid \forall p \in [I_{2 T}, \eta] \rightarrow F_{2 T}(p)<1.
    \end{align}
    So, for $p=\eta$, we have:
    \begin{align} \label{eq:16.3}
          V_{1}^2(p;q,T) > q \delta \bar{V}_{1}^2(q,T-1) + (1-q)  +  (1-q) \delta V(\bar{P}, 1, T-1).
    \end{align}
The inequality results from $p > P_{2, T}^{*} = \delta V(\bar{P}, 1, T-1)$, and the right-hand side is equal to the expected profit associated with $P_{2, T}^{*}$. In addition, for the points in a small right neighbourhood of $P_{2, T}^{*}$, we have:
    \begin{align}
        \lim _{p \rightarrow P_{2, T}^{*}+} V_{1}^2(p;q,T) = q \delta \bar{V}_{1}^2(q,T-1)  +  (1-q) \delta V\left(\bar{P}, 1, T-1\right),
    \end{align}
which is obtained from the fact that $p \rightarrow P_{2, T}^{*} = \delta V\left(\bar{P}, 1, T-1\right)$ and its substitution in $V_{1}^2(p;q,T)$.
Based on limit definition, there exists $\epsilon > 0$ as the expected profit of $P \in [p_{2, T}^{*}, P_{2, T}^{*}+\epsilon]$ is less than the expected profit of $p=\eta$. This means that there is a pure strategy with the expected profit greater than the equilibrium one, and based on theorem \ref{lem:4.1} this state cannot be an equilibrium. In other words based on Nash equilibrium definition, firm $1$ never offers a price in this interval, and we have:
    \begin{align}  \label{eq:5.15}
        I_{1 T}=\sup \left\{p \mid F_{1 T}(p)=0\right\}>p_{2, T}^*,
    \end{align}
and this is in contradiction with our initial assumption regarding the lower bound $I_{1 T}$.
\item Assume $U_{1 T}=U_{2 T} \geq I_{1 T} = I_{2 T}>P_{2, T}^{*}$. If the probability firm $2$ assigned to $U_{2 T}$ is not positive, for each $\epsilon>0$, there exists $\eta\in (p_{2, T}^{*},U_{2 T})$ as:
    \begin{align}
        \forall p \in(\eta, U_{2 T}] \rightarrow F_{2 T}(p)>1-\epsilon.
    \end{align}
When we consider $\epsilon=\frac{I_{1T} - P_{2 T}^{*}}{\frac{1}{2}\left(\delta V(\bar{P}, 1, T-1)+3 U_{2T}\right)}$, the expected profit of firm $1$ for $\mathrm{p} \in(\eta, U_{2 T}]$ satisfies the following inequality:
\begin{align} \label{eq:5.14}
 V_{1}^2(p;q,T)  &<  q \delta \bar{V}_{1}^2(q,T-1)\\
&+(1-q)[\delta V(\bar{P}, 1, T-1)+\frac{1}{2} \epsilon(\delta V(\bar{P}, 1, T-1))+\frac{3}{2} U_{2 T})] \notag \\
&= q \delta \bar{V}_{1}^2(q,T-1) + (1-q) I_{1 T}. \notag
\end{align}
The inequality is obtained by substituting $1-F_{2 T}(p)$, $F_{2 T}(p)-F_{2 T}(p_{-})$ and $p$ with their upper bounds, $1$, $\epsilon$ and $U_{2T}$ in $V_{1}^2(p;q,T)$, the following inequality is derived:
The last expression is obtained by substitution of $\epsilon$.
On the other hand, in a left neighbourhood of $I_{1T}$, the expected profit is:
    \begin{align} \label{eq:5.15}
        \lim _{p \rightarrow I_{1T}^{-}} V_{1}^2(p;q,T)= q \delta \bar{V}_{1}^2(q,T-1) + (1-q) I_{1T}
    \end{align}
So, in a left neighbourhood of $I_{1T}$, the expected profit is more than the expected profit of $p \in (\eta, U_{2 T}]$.
Thus, based on theorem \ref{lem:4.1}, this state is not possible in equilibrium. So, we should have $F_{2 T}\left(U_{2 T}\right)-F_{2T}\left(U_{2 T-}\right)>0$. 
In this case, since we have $U_{1T} >P_{2, T}^{*} = \delta V(\bar{P}, 1, T-1)$, the value of $V_{1}^2(p;q,T)$ in a small left neighborhood of $U_{1 T}$ is greater than its value in $U_{1 T}$. Accordingly, based on theorem \ref{lem:4.1}, the probability given to the points with expected value below the equilibrium expected value is zero, and based on this, the probability firm $1$ assigns to $U_{1 T}$ is zero. However, we proved that the probability assigned
to the upper bound should not be zero. As a result, we showed that state 4 is not possible to maintain
equilibrium.
\end{enumerate}
\end{proof}

\textbf{A.3 Proof of proposition \ref{lem:three}}
\begin{proof}
Considering lemma \ref{lem:one} and \ref{lem:two}, for a duopoly competition with binary demand and $T \geq 2$, the necessary condition for equilibrium is to have a pure strategy $I_{1 T}=$ $I_{2 T}=U_{1T}=U_{2T}$. Since we have $P_{2, T}^{*} = \delta V(\bar{P}, 1, T-1)$, it can be shown that this strategy is a Nash equilibrium. When $T=1$, this is a simple Bertrand model in which every firms offers  zero in equilibrium.
\end{proof}
\textbf{A.4 Proof of proposition \ref{lem:four}}
\begin{proof}
The proof of this proposition is similar to the proof of proposition \ref{lem:three}. 
These conditions are proved to be true for $N=2$. So, we can use inductive reasoning. Assume that the three sections of the proposition are valid for $\Gamma(N-1, q, T)$. Let us first demonstrate $I_{i N T} \geq  P_{N, T}^{*}$.
 Without loss of generality, let us first consider the following equation:
\begin{align*}
    I_{1 N T} = \min_{1 \leq i \leq N} I_{i N T}.
\end{align*}
Now, we use proof by contradiction. For this aim, we assume $I_{1 N T} < P_{N, T}^{*}$, and we examine the two resulting possibilities.
    \begin{enumerate}
    \item If $F_{-1 N T}\left(I_{1 N T}\right) < 1$, from right-continuity property of  cumulative distribution function, we have:
        \begin{align} \label{eq:5.17}
            \exists \eta \in\left(I_{1 N T}, p_{N, T}^{*}\right) \mid \forall p \in [I_{1 N T}, \eta] \rightarrow F_{-1 N T}(p)<1.
        \end{align}
    Now, consider $p \in [I_{1 N T}, \eta]$. Since we have $p < \delta V(\bar{P}, N-1, T-1) =  P_{N, T}^{*}$, if we substitute $p$ in \eqref{eq:defvn} with the upper bound $\delta V(\bar{P}, N-1, T-1)$, we have the following inequality:
         \begin{align} \label{eq:5.19}
             V_{1}^N(p;q,T) < q \delta \bar{V}_{i}^N(q,T-1) + (1-q)   \delta V(\bar{P}, N-1, T-1).
         \end{align}
    The right-hand side of \eqref{eq:5.19} is firm $1$'s expected profit if they decide to postpone their selling. So, based on theorem \ref{lem:4.1}, the probability assigned to the points with the expected profit less than the equilibrium expected profit must be zero in equilibrium, and as a result it is not possible that firm $1$ post a price in $[I_{1 N T}, \eta]$ in equilibrium. This is a contradiction, and this state is rejected.
    \item Assume $F_{-1 N T}(I_{1 N T})=1$. Because we assumed that $I_{1 N T}$ is the least lower bound, there is at least one firm, other than $1$, that posts a price equal $I_{1 N T}$ with probability one. As a result, there exists firm $i\neq 1$, such that $F_{i T N}(I_{1 N T}) - F_{i T N}(I_{1 N T}^{-}) > 0$. Since we have $I_{1 T N} < p_{N, T}^{*} = \delta V(\bar{P}, N-1, T-1)$, and $I_{i N T} = I_{1 N T}$ is the least lower bound, if we substitute $I_{1 N T}$ with its upper bound $\delta V(\bar{P}, N-1, T-1)$ in \eqref{eq:defvn}, we have:
         \begin{align} 
             V_{i}^N(p;q,T) < q \delta \bar{V}_{i}^N(q,T - 1) + (1-q) \delta V(\bar{P}, N-1, T-1).
         \end{align}
    The right-hand side is the expected value of firm $i$'s profit when they postpone selling. So, firm $i$ should not assign positive probability to $I_{i N T}$, and this is a contradiction.
    \end{enumerate}
Now, the first part of proposition is proved.
We again use proof by contradiction to prove the second part $\gamma_{N T}\left(p_{N, T}^{*}\right) \geq 2$. We first assume:
    \begin{align} \label{eq:5.20}
        p_{N, T}^{*} \leq I_{1 N T} < I_{2 N T} \leq \cdots \leq I_{N N T}
    \end{align}
Considering this assumption, we have:
    \begin{align} \label{eq:5.21}
        \exists \eta \in\left(I_{1 N T}, I_{2 N T}\right) \mid F_{1 N T}(\eta)>0.
    \end{align}
We only need to calculate firm $1$'s expected profit for $p \in[I_{1 N T}, \eta]$. Since firm $1$ offers the lowest price in the market, it can sell for sure if demand is not zero. So, we have:
    \begin{align} \label{eq:5.22}
        V_{i}^N(p;q,T) = q \delta \bar{V}_{i}^N(q,T - 1) + (1-q) p.
    \end{align}
The right-hand side of \eqref{eq:5.22} is bounded from above with its value for $p = \eta$. As a result, offering a price equal to $ p \in\left[I_{1 T N}, \eta\right)$ is not possible in equilibrium, and it is in contradiction with our assumption. As a result, the following relation stands:
    \begin{align}  \label{eq:5.24}
        P_{N, T}^{*} \leq I_{1 N T}=I_{2 N T} \leq \cdots \leq I_{N N T}.
    \end{align}
To prove the second part, we should also show $ P_{N, T}^{*} < I_{1 N T}=I_{2 N T}$ is not possible in equilibrium. We use proof by contradiction. Let us assume that $ P_{N, T}^{*} < I_{1 N T}=I_{2 N T}$ in equilibrium. Now, we examine and reject the two possible resulting states.
\begin{enumerate}
\item Assume $F_{-1 N T}(I_{1 N T})>0$. Then, Since $I_{1NT}$ is greater than $\delta V(\bar{P}, N-1, T-1)$, firm $1$'s expected profit satisfies the following equation:
     \begin{align} \label{eq:5.257}
         V_{1}^N(I_{1 N T};q,T) < q \delta \bar{V}_{1}^N(q,T - 1) + (1-q) I_{1 N T}.
     \end{align}
We also have the following inequality and equality for the limit of $V_{1}^N(p;q,T)$ when  $p$ approaches to $I_{1 N T}$ from its right and left neighbourhoods, respectively:   
    \begin{align}  \label{eq:5.26}
        \lim _{p \rightarrow J_{1 N T}+} V_{1}^N(p;q,T) < q \delta \bar{V}_{1}^N(q,T - 1) + (1-q) I_{1 N T},
    \end{align}
    \begin{align}  \label{eq:5.27}
        \lim _{p \rightarrow I_{1 N T}-} V_{1}^N(p;q,T) = q \delta \bar{V}_{1}^N(q,T - 1) + (1-q) I_{1 N T}.
    \end{align}
Based on the limit definition, we can find $\epsilon_{1}$ and $\epsilon_{2}$ as the expected profit of $p \in[I_{1 N T}, I_{1 N T}+\epsilon_{1})$ is less than the expected profit of $p \in\left(I_{1 N T}-\varepsilon_{2}, I_{1 N T}\right)$. So, based on theorem \ref{lem:4.1}, firm $1$ never offers a price from $[I_{1 N T}, I_{1 N T}+\epsilon_{1})$ in equilibrium, and we can reject the assumption $F_{-1}\left(I_{1 N T}\right)>0$.
\item Assume $F_{-1NT}\left(I_{1 N T}\right)=0$.
 In this situation, every firm has the option to sell at $I_{1NT}$ when there is a positive demand. So, since $I_{1 N T} > V_{1}^N(p;q,T) = P_{N, T}^{*} $, the expected profit of firm $i$, $V_{1}^N(p;q,T)$, for $p = I_{1NT}$ is strictly greater than its corresponding value for selling postponement. Thus, there is no firm tend to let others sell, and the following expression holds:
    \begin{align}  \label{eq:5.28}
          U_{1 N T}=U_{2 N T}=& \cdots=U_{N N T}. 
    \end{align}
In addition, the probability assigned to the upper bound is zero for all firms. In order to show this we examine two different possibilities. First, if firm $i$ assigns a positive probability to $U_{i N T}$, and there exists at least one another firm $j$ who assigns zero probability to $U_{j N T}$, firm $i$'s expected profit for $U_{i N T}$ is equal to the expected profit of postponement. So, from theorem \ref{lem:4.1} the probability firm $i$ assigns to $U_{i N T}$ should be zero. This is a contradiction.\\
Secondly, If all of the firms assign positive probability to the upper bound, since $\delta V(\bar{P}, N-1, T-1) < U_{i N T}$, the expected profit of firm $i$ at $U_{i N T}$ is less than its corresponding value for a small left neighborhood of $U_{i N T}$. This is because in the left neighborhood, it has more chance to sell. So, it has to assign zero probability to $U_{i N T}$, and this is a contradiction. So, we showed that the upper bounds are equal, and the probability assigned to them is zero. As a result, the expected value of profit for firm $1$ satisfies the following equation: 
    \begin{align}  \label{eq:5.29}
        \lim _{p \rightarrow U_{1 N T}{ }^{-}}V_{1}^N(p;q,T) = q \delta \bar{V}_{1}^N(q,T - 1) + (1-q)\delta V(\bar{P}, N-1, T-1).
    \end{align}
Since we have $\delta V(\bar{P}, N-1, T-1) < I_{1 N T}$, we also have:
    \begin{align}  \label{eq:5.29prim}
        \lim _{p \rightarrow U_{1 N T}{ }^{-}}V_{1}^N(p;q,T) < q \delta \bar{V}_{1}^N(q,T - 1) + (1-q) I_{1 N T}.
    \end{align}
The right-hand side is the expected profit of firm $1$ for $p =  I_{1 N T}$. So, the probability assigned to a left neighborhood of $U_{1 N T}$ must be zero, and this is a contradiction with upper bound definition.
So, we have the following result is established:
    \begin{align} \label{eq:5.30}
        I_{1 N T}=I_{2 N T}=P_{N, T}^{*}
    \end{align}
\end{enumerate}
Applying proof by contradiction, we show that at least two firms assign probability one to $P_{N, T}^{*}$. 
Assume that non of the firms assign probability one to $P_{N, T}^{*}$. First, we have that the limit of the expected profit of firm $1$ when $p$ approaches to $I_{1 N T} = P_{N, T}^{*}$ from its right neighborhood is equal to its corresponding value at $P_{N, T}^{*}$. In other words, we have:
    \begin{align} \label{eq:5.31}
        \lim _{p \rightarrow I_{1 N T}^{+}} V_{1}^N(p;q,T) =  q \delta \bar{V}_{1}^N(q,T -1) + (1-q) P_{N, T}^{*}.
    \end{align}
Secondly, we can find $\eta$ such that:
    \begin{align} \label{eq:5.32}
        \exists \eta > p_{N, T}^{*} \mid 0 < F_{-1 N T}(\eta)<1.
    \end{align}
Now, because $P_{N, T}^{*} =  \delta V(\bar{P}, N-1, T-1) < \eta$, firm $1$'s expected profit satisfies the following inequality:
\begin{align}  \label{eq:5.33}
V_{1}^N(\eta;q,T) >  q \delta \bar{V}_{1}^N(q,T - 1) + (1-q) P_{N, T}^{*}.
\end{align}
As a result of \eqref{eq:5.32} and \eqref{eq:5.33} and based on the limit definition, parameter $\epsilon$ and interval $\left[I_{1 N T}, I_{1 N T} + \epsilon\right)$ can be found such that for each price in this interval the expected profit of firm $1$ is less than the expected profit of $\eta$. So, this state is not possible in equilibrium. Therefore, we should have $F_{-i}(P_{N, T}^{*}) = 1$. Since we assume that the mixed strategy randomizations are independent in definition of mixed strategy \cite{osborne1994course}, There should be a firm $j$ that offers $I_{1 N T}$ with probability one.  With the same reasoning for firm $j$, there should be another firm $i$ that posts $I_{1 N T}$ with probability one, and the result is established. Therefore, the second part of equation is proved.\\
We can infer from the preceding sections that the expected profit is equal to $V(\bar{P}, N, T)$. Let us first prove it for $T = N$. We have $P_{N, N-1}^{*} = V(\bar{P}, N, N-1) = 0$, and based on the previous sections two firms offer zero price. So, we have:
 \begin{align} \label{eq:defvn2}
        V_{i}^N(P_{N, T}^{*};q,N ) = q \delta 0
        +(1-q) \delta V(\bar{P}, N-1, T-1) = V(\bar{P}, N, T).
 \end{align}
So, the induction reasoning can be applied. Note that there are two induction assumptions we are making to prove the third part of proposition. First, we start with the assumption we made regarding a game with $N-1$ firms, and that is applied in all the three sections' proofs. In the meantime, we made another assumption regarding a game with the planning horizon $T-1$ to prove the third section of the proposition.
As a result, we now assume that the third statement is valid for $T-1$, and we calculate the expected profit for $\Gamma(N,T,q)$ as follows:
 \begin{align} \label{eq:defvn2}
        V_{i}^N(P_{N, T}^{*};q,T ) = q \delta V(\bar{P}, N, T-1)
        +(1-q) \delta V(\bar{P}, N-1, T-1)
 \end{align}
Thus, the result is established.
\end{proof}

\textbf{A.5 Proof of proposition \ref{lem:five} }
\begin{proof}
Let us prove this by induction.
First, we must establish that $W_{1}(p)$ leads to equilibrium, and the equilibrium profit equals $V(\bar{p}, 2,1)$.
 Substituting $V(\bar{p}, 1, 0) = 0$ in the given formula for $W_{1}(p)$, we have:
    \begin{align}  \label{eq:5.35}
        W_{1}(p)=\left\{
            \begin{array}{cc}
            0 & p \leq P_{N, T}^{*} \\
            \frac{ \sum_{i=1}^{+\infty} q_{i} p - V(\bar{p}, 2, 1)}{q_{1}p} & P_{N, T}^{*}<p<\bar{p}\\
            1 & p \geq \bar{p}
        \end{array}\right.
    \end{align}
Then, if we substitute $F_{2 T}(p) = w_1(p)$ in \eqref{eq:DuopolyGEXP}, we have:
    \begin{align} \label{eq:DuopolyGEXP2}
         W_{1}^2(p;q,1)&= q_{1}p[ \frac{ V(\bar{p}, 2, 1) +q_{1}(p) - \sum_{i=1}^{+\infty} q_{i} p}{q_{1}(p)}] 
        +\sum_{i=2}^{+\infty} q_{i}p = V(\bar{p}, 2, 1)
    \end{align}
for $p \in (P_{2,1}^{*}, \bar{p}))$, and $W_{1}^2(p;q,1) = 0$ for other values of $p$. The expected value of profit for $p \in (P_{2,1}^{*}, \bar{p}))$ is similar. In conclusion, no pure strategy has an expected profit greater than the equilibrium expected profit, and there is no probability ascribed to any pure strategy that has an expected profit that is less than the equilibrium profit. As a result, the necessary and sufficient conditions of equilibrium stated in \ref{lem:4.1} are satisfied. \\
To use inductive reasoning, we assume that the conditions are satisfied for $T-1$.
substituting $W_{T}(p)$ instead of $F_{2 T}(p)$ in $W_{1}^2(p;q,T)$, we have:
\begin{align} \label{eq:5.34}
W_{1}^2(p;q,T) &= q_{0} \delta V(\bar{p}, 2, T-1) + q_{1}[ \frac{V(\bar{p}, 2, T) - q_{0} \delta V(\bar{p}, 2, T-1)-\sum_{i=1}^{+\infty} q_{i} p}{q_{1} (\delta V(\bar{p}, 1, T-1)-p)} \delta V(\bar{p}, 1, T-1) \\ & + (\frac{q_{1} (\delta V(\bar{p}, 1, T-1)-p) - V(\bar{p}, 2, T) + q_{0} \delta V(\bar{p}, 2, T-1) + \sum_{i=1}^{+\infty} q_{i} p}{q_{1} (\delta V(\bar{p}, 1, T-1)-p)} ) p] \\
&+\sum_{i=2}^{+\infty} q_{i} p
=V(\bar{p}, 2, T) \notag
\end{align}
for $p \in (P_{2,T}^{*}, \bar{p})$. 
In addition, based on the formulation of $W_{T}(p)$, we have $W_{1}^2(p;q,T) = 0$ for $p \notin (P_{2,T}^{*}, \bar{p})$. As a result, the necessary and sufficient of equilibrium in \ref{lem:4.1} are satisfied, and the result is established.
\end{proof}
\textbf{A.6 Proof of lemma \ref{lem:six}.}
Let's start by demonstrating that $I_{1 T} = I_{2 T}$. To employ proof by contradiction, assume $I_{1T} < I_{2T}$. According to this assumption, there exists $\eta > I_{1T}$ such that $F_{1T}(\eta_{-})>0$. In this case, since firm $2$ is absent, firm $1$ acts like a monopolist. Therefore, the expected profit of $p \in [I_{1T},\eta)$ is smaller than the expected value resulting from offering $\eta$. This interval should not have a positive probability in equilibrium according to the theorem \ref{lem:4.1}, which is a contradiction. So, we have $I_{1 T}=$ $I_{2 T}$ in equilibrium. Now, we need to show $I_{1T} = I_{2T} = P_{2T}$. We again use proof by contradiction. 
If we assume $I_{1T} = I_{2T} < P_{2T}$, there will be two possible states. Here, we examine both of them and demonstrate why it is impossible for them to establish an equilibrium.
\begin{enumerate}
    \item $F_{2 T}\left(I_{2 T}\right)<1$: Because cumulative distribution function is right-continues, we have:
    \begin{align}  \label{eq:5.37}
            \exists \eta \in (I_{2 T}, P_{2, T}^{*}) \mid \forall p \in[I_{2 T}, \eta) \rightarrow F_{2 T}\left(I_{2 T}\right) \leq F_{2 T}(p)<1.
    \end{align}
Since when $\sum_{1=2}^{+\infty} q_{i}>0$, we have $ P_{2 T}^{*} > \delta V(\bar{P}, 1, T-1)$, we can substitute $P_{2 T}^{*}$ as an upper bound for both $p$ and $\delta V(\bar{P}, 1, T-1)$ in $W_{1}^2(p;q,T)$. As a result of this substitution, we obtain the following inequality:
    \begin{align}  \label{eq:5.39}
            W_{1}^2(p;q,T) < q_0 \delta \bar{W}_{1}^2(q,T-1) + q_{1} P_{2 T}^{*} +\sum_{i=2}^{+\infty} q_{i} P_{2 T}^{*}.
    \end{align}
for $p \in[I_{2 T}, \eta)$. From \eqref{eq:optionofgen} and \eqref{eq:optionofgen2}, we have:
\begin{align} \label{eq:pbarpstar}
    (1-q_{0}) P_{N, T}^{*} = q_{1} \delta V(\bar{p}, 1, T-1)+\sum_{i=2}^{+\infty} q_{i} \bar{p}.
\end{align}
Now, we substitute $ (1-q_{0}) P_{N, T}^{*} $ in \eqref{eq:5.39}:
    \begin{align}  \label{eq:5.39prim}
        W_{1}^2(p;q,T) <  q_0 \delta \bar{W}_{1}^2(q,T-1)  +  q_{1} \delta V(\bar{P}, 1, T-1)  + \sum_{i=2}^{+\infty} q_{i} \bar{P}.
    \end{align}
The right-hand side of \eqref{eq:5.39} demonstrates the expected value associated with offering the maximum price $\bar{P}.$ So, the expected benefit of $p \in[I_{2 T}, \eta)$ is less than its value for $p = \bar{P}$. So, we found a pure strategy with expected profit greater than the equilibrium expected profit, and this stat cannot establish an equilibrium based on theorem \ref{lem:4.1}.
\item Assume $F_{2T}(I_{2 T})=1$. The intuition is comparable to the prior section.
Because $P_{2, T}^{*} > \delta V(\bar{P}, 1, T-1)$ and by \eqref{eq:optionofgen}, firm $2$ achieves less profit for $p = I_{2T}$ than $p = \bar{P}$: 
    \begin{align} \label{eq:5.40}
        W_{2}^2(p;q,T)
        <  q_0 \delta \bar{W}_{2}^2(q,T-1)  + q_{1} P_{2 T}^{*}+\sum_{i=2}^{+\infty} q_{i} P_{2 T}^{*}
    \end{align}
 So, refusing these two states, we have $I_{1T}=I_{2T} \geq P_{2T}^{*}$ in equilibrium.
\end{enumerate}
Now, we only need to show $p_{N, T}^{*}<     I_{1 T}=I_{2 T}$ is not possible to establish an equilibrium. We again employ proof by contradiction. Below, we examine the two possible states arising from our assumption $p_{N, T}^{*}< I_{1 T}=I_{2 T}$.
\begin{enumerate}
    \item  Assume $F_{2 T}(I_{2 T}) > 0$. Then, because we have $I_{1 T}>p_{N, T}^{*}$ and $p_{N, T}^{*}>\delta V(\bar{P}, 1, T-1)$, substituting $I_{1 T}$ as an upper bound instead of $\delta V(\bar{P}, 1, T-1)$ in $W_{1}^2(I_{1 T};q,T)$, we have: 
        \begin{align} \label{eq:5.41}
             W_{1}^2(I_{1 T};q,T) < q_{0} \delta \bar{W}_{2}^2(q,T-1) + q_{1} I_{1 T} + \sum_{i=2}^{+\infty} q_{i} I_{1 T}.
         \end{align}
    By same intuition, when $p$ approaches to $I_{1T}$ from its right neighbourhood, we have the following inequality:
        \begin{align} \label{eq:5.44}
            \lim _{P \rightarrow I_{1 T^{+}}} W_{1}^2(q;T) < q_{0} \delta \bar{W}_{2 } ^2(q,T-1) + q_{1} I_{1 T} + \sum_{i=2}^{+\infty} q_{i} I_{1 T}
        \end{align}
    On the other hand, when $p$ approaches to $I_{1T}$ from its left neighbourhood, the limit of expected profit is calculated as follows:
        \begin{align} \label{eq:5.45}
             \lim _{P \rightarrow I_{1 T^{-}}}W_{1}^2(q;T) =q_{0} \delta V_{1}(T-1)+q_{1} I_{1 T}+\sum_{i=2}^{+\infty} q_{i} I_{1T}.
        \end{align}
    Based on limit definition, there are $\epsilon_{1}>0$ and $\epsilon_{2}>0$ such that for $p \in [I_{1T}, I_{1 T} + \epsilon_{1})$ the expected profit is less than the expected profit for $ p \in (I_{1 T}-\varepsilon_{2}, I_{1 T})$. According to theorem \ref{lem:4.1}, firm $1$ never assign a positive probability to this interval in equilibrium, and this is a contradiction.
    \item Assume $F_{2T}\left(I_{2T}\right)=0$. as a result of the prior state being rejected, we also have $F_{1T}\left(I_{1T}\right)=0$. In this situation, both firms have an option to sell for $I_{1T}$. Based on our assumption $p_{N, T}^{*} < I_{1 T} = I_{2 T}$, and the relationship we have in \eqref{eq:pbarpstar}, the expected profit of postponing the sell is less than the existing option of offering $I_{1 T} = I_{2 T}$. Therefore, probability assigned to postponement is zero, and the following equation holds:
        \begin{align} \label{eq:5.46}
        U_{1 T} =U_{2 T} \leq \bar{P}.
        \end{align}
    In addition, it can be shown that the probability assigned to upper bounds cannot be positive.
    This is due to the fact that if only one of them gives the upper bound a positive probability, that would be the same as giving the postponement a positive probability, which we have shown is impossible in equilibrium.
    Furthermore, it can be demonstrated that the expected profit of the upper bounds is lower than the expected profit associated with the prices in a left neighborhood of the upper bound, making it impossible for both of them to simultaneously assign a positive probability to the upper bound.  
    The reason is that $V(\bar{P}, 1, T-1) < p_{N, T}^{*} < I_{1 T}=I_{2 T} \leq U_{1 T} = U_{2 T}$.\\
    Therefore, when $p$ approaches to $U_{1T}$ from its left neighbourhood, the limit of firm $1$'s expected profit's  can be obtained as:
    \begin{align} \label{eq:5.48}
        \lim _{P \rightarrow U_{1 T^{-}}} W_{1}^2(p;q,T) = q_{0} \delta \bar{W}_{2 } ^2(q,T-1) + q_{1} \delta V(\bar{P}, 1, T-1) + \sum_{i=2}^{+\infty} q_{i} U_{1 T}.
    \end{align}
    Considering \eqref{eq:optionofgen}, \eqref{eq:optionofgen2} and our assumption $p_{N, T}^{*} < I_{1 T} = I_{2 T}$, firm $1$'s expected  profit for $I_{1 T}$ is more than for a right neighborhood of $U_{1 T}$:
        \begin{align} \label{eq:5.49}
            \lim _{p \rightarrow U_{1 T^{-}}} W_{1}^2(p;q,T) &\leq q_{0} \delta \bar{W}_{2 } ^2(q,T-1) + (1-q_{0}) P_{N, T}^{*} \\
            & < q_{0} \bar{W}_{2 } ^2(q,T-1) + (1-q_{0}) I_{1 T}.
        \end{align}
    Based on limit definition, there exists $\epsilon_{1}>0$ such that for $ p \in\left(U_{1 T}-\varepsilon_{1}, U_{1 T}\right)$, the expected profit is less than the expected profit for $I_{1T}$. So, firm $1$ never assign a positive probability to this interval in equilibrium, and it is a contradiction. Thus, the following equation is proved: 
        \begin{align} \label{eq:5.50}
            l_{1 T} = I_{2 T} = P_{2, T}^{*}.
        \end{align}
    \end{enumerate}
Now, we examine the second part of lemma \ref{lem:six}.
First, assume $U_{1T} \neq U_{2 T}$ and $U_{1 T} < U_{2T}$. According to the previous section, we have $ p_{2,T}^{*} \leq U_{1T} < U_{2T}$.
So, we have:
\begin{align} \label{eq:5.51}
    \exists \eta \in\left(U_{1 T}, U_{2 T}\right) \mid 0 < F_{2 T}(\eta) < 1.
\end{align}
Firm $2$'s expected profit can be calculated from the following equation for $p \in (U_{1T},U_{2T} ]$:
\begin{align} \label{eq:5.52}
   \begin{aligned}
        W_{2}^2(p;q,T)=q_{0} \delta V_{1}(T-1)+q_{1} \delta \bar{W}_{2 } ^2(q,T-1) +\sum_{l=2}^{+\infty} q_{i} p \\
        < q_{0} \delta V_{1}(T-1)+q_{1}& \delta V(\bar{P}, 1, T-1)+\sum_{l=2}^{+\infty} q_{l} U_{2 T}
    \end{aligned}
\end{align}
The right-hand side of \eqref{eq:5.52} is the expected profit at $U_{2T}$. As a result, since firm $1$ is absent in $p \in (U_{1T},U_{2T} ]$, firm $2$ assigns no probability to $(U_{1T},U_{2T})$. In addition, because we know $U_{2T}$ is greater than $U_{1T}$, firm $2$ has to  assigns positive probability to the upper bound $U_{2T}$. 
In addition, it is not possible to have equilibrium if firm $2$ assigns a positive probability to $U_{1T}$. To see this, we use proof by contradiction. 
Thus, assuming $F_{2T}(U_{1T})- F_{2T}(U_{1T-})>0$, firm $1$'s expected profit for $U_{1T}$ is less than its value for a right neighborhood of $U_{1T}$. This is due the fact that we have $V(\bar{P}, 1, T-1) < p_{N, T}^{*} < U_{1T}$. 
So, firm $1$ assigns zero probability to $U_{1T}$.  Considering all we've mentioned, firm $2$'s expected profit of $U_{1T}$ would be less than its corresponding value for $U_{2T}$. So, we showed that firm $2$ assigns zero probability to $U_{1T}$. Let's examine at the situation this time from the standpoint of firm $1$. Because firm $2$ is absent in $(U_{1T},U_{2T})$, we have:
        \begin{align} \label{eq:5.53}
            W_{1}^2(U_{1 T};q,T) = \lim _{P \rightarrow U_{1 T^{-}}} W_{1}^2(p;q,T) = q_{0} \delta \bar{W}_{1} ^2(q,T-1)
            + q_{1}[F_{2 T}(U_{1 T}) \delta V(\bar{P}, 1, T-1)
           \\ +(1-F_{2 T}(U_{1 T})) U_{1 T}] 
            +\sum_{i=2}^{+\infty} q_{i} U_{1 T} < q_{0} \delta \bar{W}_{1} ^2(q,T-1)
            + q_{1}[F_{2 T}(U_{1 T}) \delta V(\bar{P}, 1, T-1) \notag
           \\ +(1-F_{2 T}(U_{1 T})) \eta] 
            +\sum_{i=2}^{+\infty} q_{i} \eta.\notag
       \end{align}
when $\eta \in (U_{1T},U_{2T})$. The right-hand side of \eqref{eq:5.53} demonstrates the expected profit of $\eta$.
Thus, we showed that there is a pure strategy for firm $1$ for which the expected profit is greater than the expected profit of the prices in a left neighborhood of $U_{1T}$. Based on theorem \ref{lem:4.1}, firm $1$ should assign zero probability to an interval $(U_{1T} - \epsilon,U_{1T}]$, when $\epsilon$ is small positive value. This is a contradiction, and we have $U_{1 T}=U_{2 T}$ in equilibrium.\\
In order to prove that $U_{1 T}=U_{2 T}<\bar{P}$ is not feasible, we employ proof by contradiction. We assume $U_{1 T} = U_{2 T} < \bar{P}$ in equilibrium. According to this assumption, there are two probable states that we examine here.
\begin{enumerate}
    \item Assume $1 - F_{1 T}(U_{1 T-})>0$. The limit of the expected profit of firm $2$, when $p$ approaches to $U_{1 T}$ from its left neighbourhood is calculated as follows:
        \begin{align} \label{eq:5.53}
            \lim _{P \rightarrow U_{1 T^{-}}} W_{2}^2(p;q,T) = q_{0} \delta \bar{W}_{2}^2(q,T-1) 
            + q_{1}[F_{1 T}(U_{1 T_{-}}) \delta V(\bar{P}, 1, T-1)
           \\ +(1-F_{1 T}(U_{1 T_{-}})) U_{1 T}] 
            +\sum_{i=2}^{+\infty} q_{i} U_{1 T}.\notag
       \end{align}
     We also can calculate the expected value of profit at $U_{1 T}$ as follows:
       \begin{align} \label{eq:5.53prim}
            W_{2}^2(U_{1 T};q,T) &= q_{0} \delta \bar{W}_{2}^2(q,T-1) 
            + q_{1}[F_{1 T}(U_{1 T_{-}}) \delta V(\bar{P}, 1, T-1)
             \\&+ (1 -F_{1 T}(U_{1 T_{-}})) \frac{1}{2}(V(\bar{P},1, T-1)+U_{1 T}) 
            +\sum_{i=2}^{+\infty} q_{i} U_{1 T}.\notag
       \end{align}
Comparing the right-hand side of \eqref{eq:5.53} and \eqref{eq:5.53prim}, we have:
\begin{align} \label{eq:5.54}
\lim _{P \rightarrow U_{2 T}-} W_{2}^2(p;q,T) > W_{2}^2(U_{2 T};q,T).
\end{align}
Relationship \eqref{eq:5.54} holds because $\delta V(\bar{P}, 1, T-1) < P_{2, T}^{*}$ and $P_{2,T}^{*} \leq U_{1 T}$, and $W_{2}^2(p;q,T)$ in the left neighbourhood of $U_{1 T}$ assigns less probability to $\delta V(\bar{P}, 1, T-1)$ than in $U_{1 T}$. 
Therefore, firm $2$ would not give a positive probability to $U_{2 T}$.
As a result, firm $1$ gains less profit at $p=U_{1 T}$ than $p=\bar{P}$, because both $p=U_{1 T}$ and $p=\bar{P}$ can be effective only when there are more than two units of demand in the market. This is in contradiction with equilibrium conditions stated in theorem \ref{lem:4.1}. So, this state is not possible in equilibrium.
    \item If $F_{iT}\left(U_{iT}\right)-F_{i T}\left(U_{iT-}\right)=0$ for $i=1$ and $2$, it can be proved that equilibrium is not possible. When $p$ approaches to $U_{1T}$ from its left neighbourhood, based on limit definition, there is $\epsilon>0$ as for $p \in(U_{1 T}-\epsilon, U_{1 T}]$ the expected profit is less than the expected profit at $p=\bar{P}$, and this is in contradiction with our assumption. 
    \end{enumerate}
So, we showed that $U_{1 T} = U_{2 T}=U \geq \bar{P}$. Also, it is not possible to offer a price greater than $\bar{P}$ in equilibrium. This is because offering this price will not result in sale when there is extra demand, and it results in a profit less than when they offer $\bar{P}$. Thus, the result is established, and we have $U_{1 T} = U_{2 T}=\bar{P}$.

\textbf{A.7 Proof of lemma  \ref{lem:seven}}

If the domain of two firms' strategies is $[P_{2, T}^{*}, \bar{P}]$, both firms assign zero probability to $P_{2,T}^{*}$. To show this, we can employ proof by contradiction. Assume firm $2$ assigns a positive probability to $P_{2,T}^{*}$. Then, we have: 
    \begin{align} \label{eq:5.56}
           \lim _{p \rightarrow P_{2, T}^{*}-}  W_{2}^2(p;q,T) = q_{0} \delta \bar{W}_{i}^2(q,T-1) 
            +q_{1} P_{2, T}^{*} + \sum_{i=2}^{+\infty} q_{i} P_{2, T}^{*},\\
            \lim_{\substack{p \rightarrow P_{2 ,T}^{*}+}} W_{2}^2(p;q,T) =  W_{2}^2( P_{2 ,T}^{*};q,T) < q_{0} \delta \bar{W}_{i}^2(q,T-1)  + q_{1} P_{2,T}^{*} + \sum_{T=2}^{+\infty} q_{i} P_{2 T}^{*}.\label{eq:5.57}
    \end{align}

We have \eqref{eq:5.57}, due to the fact that $P_{2 , T}^{*}>\delta V(\bar{P}, 1, T-1)$. As a result, based on limit definition, there are $\epsilon_{1}>0$ and $\epsilon_{2}>0$ such that for $ p \in [P_{2, T}^{*}, P_{2, T}^{*}+\varepsilon_{1})$, the expected profit is less than the expected profit for $ p \in\left(P_{2 ,T}^{*}-\epsilon_{2}, P_{2 T}^{*}\right)$. This is a contradiction with our assumption regarding $P_{2 ,T}^{*}$ being the lower bound. So, both firms assign a probability equal to zero to $P_{2 ,T}^{*}$.
Based on theorem \ref{lem:4.1}, the expected profit of a set of pure strategies cannot be grater than the equilibrium expected profit. We also have that the expected profit of a set of pure strategies with positive assigned probability should not be less than the equilibrium expected profit. Furthermore, we know that there $\forall \epsilon>0$, we have $F_{1T}(P_{2 T}^{*} + \epsilon )>0$. For a small $\epsilon$, we assume the expected profit of the set of points in $(P_{2 T}^{*}, P_{2 T}^{*} + \epsilon)$ is equal to a common value $W$. Since we have: 
    \begin{align} \label{eq:5.56}
          W = \lim_{\substack{p \rightarrow P_{2 ,T}^{*}+}} W_{2}^2(p;q,T) = q_{0} \delta \bar{W}_{i}^2(q,T-1)  + q_{1} P_{2,T}^{*} + \sum_{T=2}^{+\infty} q_{i} P_{2 T}^{*},
    \end{align}
and the initial condition $\bar{W}_{i}^2(q,1) = V(\bar{P}, 2, 1) =0$, we can conclude from \eqref{eq:optionofgen2}, that the expected profit in equilibrium is $V(\bar{P}, 2, T)$, which is similar for both firms. \\
To complete the proof, we again use proof by contradiction.
Let us assume that the equilibrium strategy is not symmetric, then we have:
\begin{align} \label{eq:5.59}
    \exists p \in (P_{2 T}^{*}, \bar{P}) \rightarrow F_{2 T}(p)>F_{1 T}(p).
\end{align}
Because cumulative distribution function is right-continues, for any small parameters $\epsilon_{1}, \epsilon_2>0$ there exists $\eta$ such that:
\begin{align} \label{eq:5.60}
    F_{i T}(\eta)-F_{i T}\left(\eta^{-}\right)=0
    \end{align}
    \begin{align} \label{eq:5.61}
    F_{i T}(\eta)-F_{i T}(p) = e_i <\epsilon_{i}.
\end{align}
Then, firm $1$'s and firm $2$'s expected profit at $p=\eta$ can be calculated from the following equations:
\begin{align} \label{eq:5.62}
 W_{1}^2(\eta;q,T) = q_{0} \delta \bar{W}_{1}^2(q,T-1) &+ q_{1}[(F_{2 T}(p)+e_{2}) \delta V(\bar{P}, 1, T-1)  \\
 &+(1-\left(F_{2 T}(p)+e_{2})) \eta\right] 
+\sum_{i=2}^{+\infty} q_{i} \eta \notag
\end{align}
\begin{align} \label{eq:5.63}
W_{2}^2(\eta;q,T) = q_{0} \delta \bar{W}_{2}^2(q,T-1) &+ q_{1}[(F_{1 T}(p)+e_{1}) \delta V(\bar{P}, 1, T-1) \\
&+ (1-(F_{1 T}(p)+e_{1})) \eta] 
+\sum_{i=2}^{+\infty} q_{i} \eta. \notag
\end{align}
Considering negligible $\epsilon_{1}, \epsilon_2$, firm $1$'s expected profit is greater than firm $2$'s in $\eta$. This is in contradiction with our previous result of similar equilibrium profit. So, the assumption of different equilibrium distribution function cannot be true, and the result is established.

\textbf{A.8 Proof of proposition \ref{lem:eight}}.
This can be shown regarding the result of proposition \ref{lem:five} and lemma \ref{lem:eight}. 

\textbf{A.9 Proof of proposition \ref{lem:11}.}
To show that is a mixed-strategy equilibrium, we have to show that it satisfies the necessary and sufficient conditions in theorem \ref{lem:4.1}. In addition, we need to show that it is a cumulative distribution function, so it should be a monotonically increasing function on $[0,1]$.
We have the following equation according to the definition:
    \begin{align}  \label{eq:5.66}
        V(\bar{P}, N, 0)=0.
    \end{align}
In addition, if we assume that all the firms have a common mixed-strategy $F(p)$,  the expected profit of firm $1$ can be calculated as follows:
    \begin{align}  \label{eq:5.671}
        W_{1}^N(p;q,1) = q_{0} \delta \bar{W_1^N(q,0)} + \sum_{i = 1}^{N-1} q_{i} Z_{i-1 , N}(F(p)) p 
        &+\sum_{i=N}^{+\infty} q_{i} p
    \end{align}
Now, we obtain the equilibrium strategy which results in $W_{1}^N(p;q,1) = V(\bar{P}, N, 1)$. Thus, by substituting $W_{1}^N(p;q,1) = V(\bar{P}, N, 1)$ and $\bar{W_1^N(q,0)} = V(\bar{P}, N, 0) = 0$ in \eqref{eq:5.671}, we can conclude the following equation:
    \begin{align}  \label{eq:5.672}
         \sum_{i=1}^{N-1} q_{i} Z_{i-1, N}(F(p)) p 
         &+\sum_{l=N}^{+\infty} q_{i} p =  \sum_{i=n}^{+\infty} q_{i} \bar{p}.
    \end{align}
Thus, if we solve $\eqref{eq:5.672}$ for $F(p)$, we have:
    \begin{align} \label{eq:5.68}
        &F(p)=\left\{\begin{array}{cc}
        0 & p \leq P_{N, T}^{*} \\
        G^{-1}(p) & P_{N, T}^{*}<p<\bar{p} \\
        1 & p \geq \bar{p}
        \end{array}\right. 
    \end{align}
    \begin{align} \label{eq:5.69}
        &G(x)=\frac{\sum_{i=n}^{+\infty} q_{i} \bar{p}}{\sum_{i=n}^{+\infty} q_{i}+\sum_{i=1}^{n-1} q_{i} Z_{i-1}(x)}.
    \end{align}
Since we have $W_{N, 1}(p) = F(p)$ in $\eqref{eq:5.671}$, all three equilibrium conditions are satisfied. 
Because the assumption of inductive reasoning is proved, we can assume that we have the equilibrium profit for $T-1$. As a result, $W_{1}^N(p;q,T)$ can be calculated as follows:
\begin{align} \label{eq:5.70}
     W_{1}^N(p;q,T) = q_{0} V(\bar{P},N, T-1) +\sum_{i=1}^{N-1} q_{i}[Z_{i-1, N}(F(p)) p+(1 - Z_{i-1, N}(F(p))&  \\ \delta V(\bar{p}, N-i, T-1)] 
    +\sum_{i=N}^{+\infty} q_{i} p =  \sum_{i=0}^{N-1} q_{i} \delta V(\bar{p}, N-i, T-1) + \sum_{i=N}^{+\infty} q_{i} \bar{p}.& \notag
\end{align}
If we solve \eqref{eq:5.70} for $F(p)$
, we have $F(p) =  W_{N, T}(p)$. So, the equilibrium conditions are satisfied, and the result is established.

\textbf{A.10 Proof of lemma \ref{lem:12}}.
To show that $W_{N, T}(p)$ is a function, we first should show that its inverse function is an injective function. Now we show that its inverse function is monotonous and increasing at $p \in\left(P_{N, T}^{*} \bar{p}\right)$. The denominator and numerator of the function is positive and monotonous, so because $Z_{K, N}(x)$ is a continues and strictly decreasing this function is monotonous. The derivative of these functions can be calculated from the following relations:
\begin{align} \label{eq:5.64}
    \frac{d G(x)}{d x}=\sum_{i=1}^{n-1} q_{i}^{2}(\delta V(\bar{p}, N-i, T-1)-\bar{P}) \frac{d Z_{i-1}(x)}{d x}
\end{align}
\begin{align}  \label{eq:5.65}
    \forall x \in(0,1) \rightarrow \frac{d Z_{k}(x)}{d x}=-(k+1)\left(\begin{array}{l}
    n-1 \\
    k+1
    \end{array}\right)(1-x)^{n-2-k} x^{k}<0
\end{align}
So, $W_{N, T}(p)$ is an increasing monotonous function at $p \in\left(P_{N, T}^{*}, \bar{p}\right)$.\\
\textbf{A.11 Proof of proposition \ref{lem:13}.}\\
Assuming \eqref{eq:lowerbound23}
, the probability every firm assigns to lower bound is zero. This can be shown by employing a proof by contradiction. If a firm $2$ assigns positive probability to this price, firm $1$'s expected profit when $p$ approaches to $P_{N, T}^{*}$ from its left neighbourhood is more than its expected profit at $P_{N, T}^{*}$ and more than its expected profit when $p$ approaches to $P_{N, T}^{*}$ from its right-neighbourhood. This is in contradiction with equilibrium conditions.
Here, we have the limit of the expected profit of firm $1$ as follows:
    \begin{align}  \label{eq:5.71}
        \lim_{P \rightarrow p_{N, T}^*-}  W_{1}^N(p;q,T)=\delta \bar{W}_{1}^N(q,T-1)+\sum_{i=1}^{\infty} q_{t} p_{N, T}^{*}.
    \end{align}
Since $\sum_{l=2}^{+\infty} q_{i}>0$, we have $V(\bar{P}, N-i, T-1) < P_{N, T}^{*}$. As a result, when at least one of other $N-1$ firms assign a positive probability to $P_{N, T}^{*}$, we have the following equation at $p = P_{N, T}^{*}$:
\begin{align} \label{eq:5.72}
    \lim_{p \rightarrow P_{N, T}^*+}  W_{1}^N(p;q,T)
    = W_{1}^N(p;q,T) < \lim_{P \rightarrow p_{N, T}^*-}  W_{1}^N(p;q,T)
\end{align}
Based on \eqref{eq:5.71} and \eqref{eq:5.72} and limit definition, we can find two intervals at the left and right of $p_{N, T}^*$, as the expected profit of the points in the left interval is greater than the expected profit of the points in the right interval containing $p_{N, T}^{*}$. So it is in contradiction with equilibrium definition and firm $1$ and firm $2$ assign zero probability to $p_{N, T}^*$ and the expected profit is $V(\bar{P}, N, T)$. (It can be shown with same reasoning as the one we employed for duopoly)\\
If we assume that the equilibrium strategy is asymmetric, we can find $p$ as the following equation holds:
\begin{align} \label{eq:5.74}
    F_{-1 N T}(p) > F_{-2 N T}(p).
\end{align}
Because cumulative distribution function is right-continues, $\epsilon$ and $\eta$ can be found as for $i=1,2$ the equations \eqref{eq:5.75} and \eqref{eq:5.76} hold: 
    \begin{align} \label{eq:5.75}
        &F_{-i N T}\left(\eta^{-}\right)-F_{-i N T}(\eta)=0, 
        \end{align}
        \begin{align} \label{eq:5.76}
        &F_{-i N T}(\eta)-F_{-i N T}(p) = e_i<\epsilon_{i}.
    \end{align}
The existence of $p$ can be shown from the fact that we assume the random strategies are independent, so the we have $F_{1 N T}(p)= 1 - \prod_{i=2}^N(1-F_{iNT}(p))$. So, different values for $F_{iNT}$ and $F_{jNT}$ would result in different values for $F_{-iNT}$ and $F_{-jNT}$.
Based on limit definition, $\epsilon_i$ can be chosen as small as we wish, and firm $1$ and firm $2$ 's expected profit can be obtained from the following equations:
    \begin{align} \label{eq:5.77}
        W_{1}^N(p;q,T) = \delta \bar{W}_{1}^N(q,T-1)\\
        &+\sum_{i=1}^{N-1} q_{i}\left[\left(F_{-1 N T}(p)+e_{1}\right) \delta V(\bar{P}, N-i, T-1)\right.\\
        &\left.+\left(1-\left(F_{-1 N T}(p)+e_{1}\right)\right) \eta\right]+\sum_{i=2}^{+\infty} q_{i} \eta \notag
    \end{align} \label{eq:5.78}
\begin{align}
    W_{2}^N(p;q,T)==q_{0} \delta \bar{W}_{2}^N(q,T-1) \\
    &+\sum_{i=1}^{N-1} q_{i}\left[\left(F_{-2}(p)+e_{2}\right) \delta V(\bar{P}, N-i, T-1)\right.\\
    &\left.+\left(1-\left(F_{-2}(p)+e_{2}\right)\right) \eta\right]+\sum_{i=2}^{+\infty} q_{i} \eta \notag
\end{align}
for $p \in (p,p + \min(\epsilon_1 , \epsilon_2))$
Considering a small enough $\epsilon_{i}$, firm $1$'s expected profit is less than firm $2$'s and it is in contradiction with our assumption of identical expected profit. Thus, the assumption of different equilibrium distribution function is not true.

\section{Numerical Results} \label{sec:num}
\begin{figure}
    \centering
    \includegraphics[scale = 0.5]{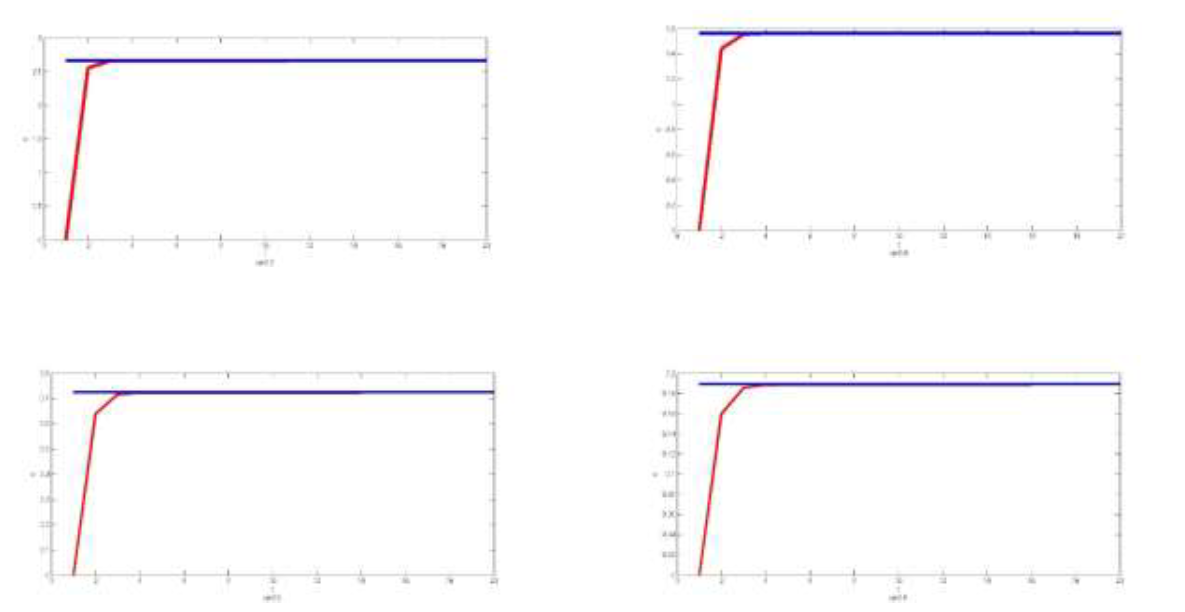}
    \caption{Duopoly model with binary demand}
    \label{fig:duo2}
\end{figure}

\begin{figure}
    \centering
    \includegraphics[scale = 0.5]{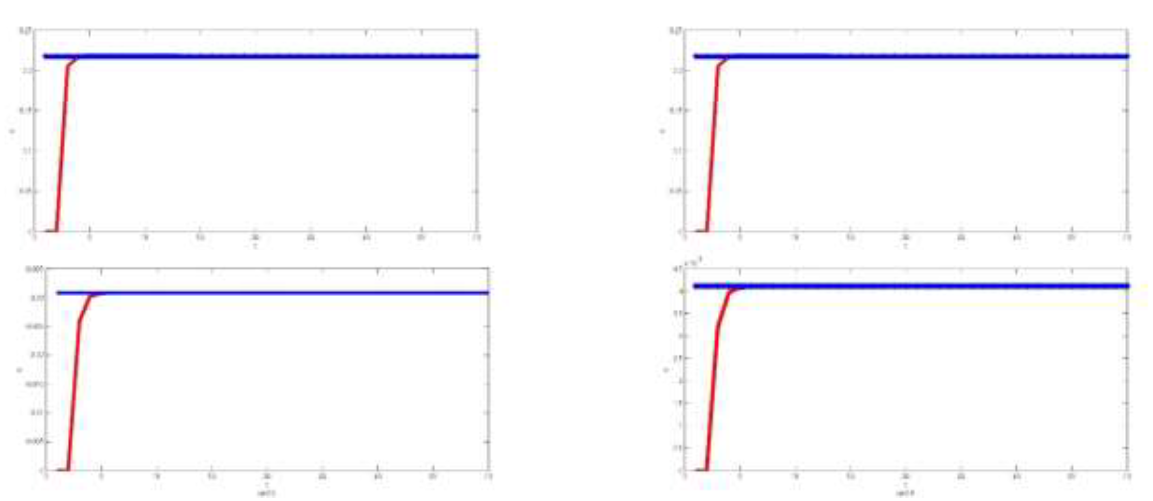}
    \caption{Opligopoly model with binary demand}
    \label{fig:olg2}
\end{figure}

\begin{figure}
    \centering
    \includegraphics[scale = 0.5]{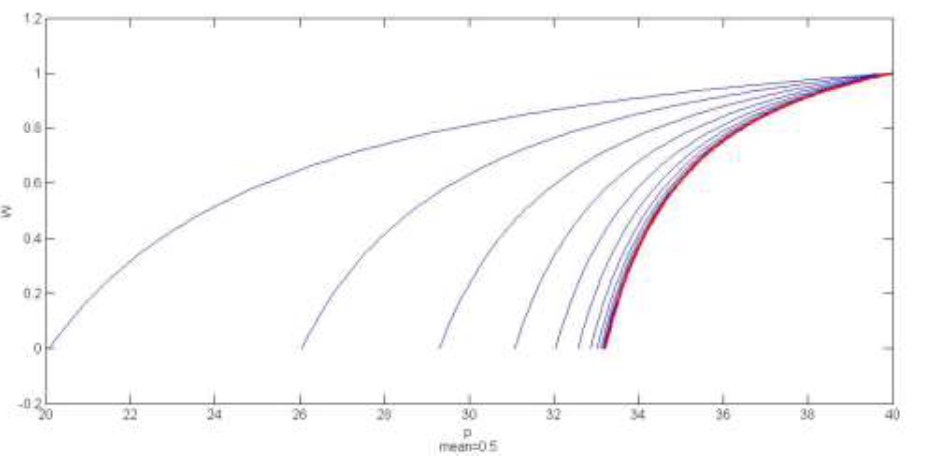}
    \caption{Duopoly model with general demand}
    \label{fig:doug}
\end{figure}

\begin{figure}
    \centering
    \includegraphics[scale = 0.5]{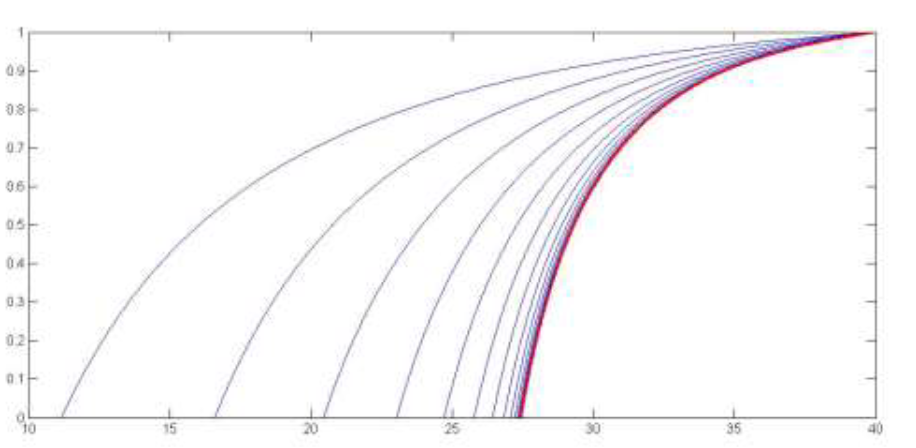}
    \caption{Oligopoly model with general demand}
    \label{fig:olgg}
    \end{figure}

  In this chapter, the results from the previous chapters are analyzed, along with the impact of various parameters.
The derived distribution functions are used to study the equilibrium strategies. Additionally, numerical examples and graphs are presented to help the reader grasp the results.
First, as part of the model evaluation and validation process, large values of the planning horizon are considered in the obtained models, and the outcomes are compared with the results of \cite{sun2017dynamic}. In these experiments, we consider $\delta = 0.9$ and $\bar{P} = 40$. \\
Here, in figure \ref{fig:duo2}, we have obtained the equilibrium price of duopoly competition from our model for $T = 2 \ldots 1000$ and the associated result from $\cite{sun2017dynamic}$. The blue lines indicate the result of \cite{sun2017dynamic} and the red lines show the result of this research. The four figure are showing the results corresponding to $q= 0.2,0.4,0.6,0.8$, respectively.
Same experiment is conducted for oligopoly model with $N = 3$, and the effective prices in equilibrium are compared in figure \ref{fig:olg2}. 
As we showed in the preceding sections, when $\sum_{i=2}{+\inf}q_i>0$, then we have a mixed-strategy equilibrium. In figure \ref{fig:doug} and \ref{fig:olgg}, we have the equilibrium distribution of our model for $N=2$ and $N=3$, respectively. The result of our model is colored in blue for $T = 2 \ldots 1000$, and the equilibrium distribution of \cite{sun2017dynamic} is denoted by red color. We considered that the demand follows a poison distribution with expected value of $0.5$.
\section{Conclusions} \label{sec:conclusion}
Four limited capacity pricing models are described in this study, and equilibrium strategies are found. The equilibrium strategies are symmetric except for the duopoly case with binary demand. 
This study show that up to
According to efficient price criteria, the price dispersion is zero while there is no chance of an excess of demand, but when that possibility arises, the market will not have a single price and will only have a predictable price pattern. 

The equilibrium strategies in this research can be used by vendors of limited capacity perishable commodities to enhance their profits.
Sellers are likely to have lower market prices that generate the option value in cases of low potential demand values, and they hardly select and offer high prices that turn into a form of risk.
The numerical data-based results indicate that as the planning horizon is increased, the results of the model with a restricted planning horizon tend to the results of the model presented by \cite{sun2017dynamic}.
Thus, when the planning horizon is large, the model in \cite{sun2017dynamic} can be a good approximation. Each vendor can only sell one product, according to the models described in this study.
Future studies can examine generalizing the model and investigating equilibrium conditions for larger capacities. Such models would be more compatible with real world conditions. 
This research did not prove the uniqueness of the equilibrium strategy in the oligopoly model with general demand. In future study, requirements for rejecting asymmetric strategies will be taken into consideration.
Also the option is a mental image in the minds of sellers before it is a mathematical formula Which is equated here with the profits from becoming a monopolist. The option might be different and the same analysis might be used with a slight adjustment depending on the circumstances in other markets.

\bibliographystyle{plain}
\bibliography{nil}

\end{document}